\documentclass[a4paper,notitlepage,12pt]{article}

\begin{document}
\title{Thermocapillary migration of a planar droplet at moderate and large Marangoni numbers}
\author{ Zuo-Bing Wu$^{*1}$ and Wen-Rui Hu$^+$\\
State Key Laboratory of Nonlinear Mechanics$^*$\\
 and National Microgravity Laboratory$^+$,\\
 Institute of Mechanics, Chinese Academy
of Sciences,\\
 Beijing 100190, China}
 \maketitle

\footnotetext[1]{Corresponding author. Tel:. +86-10-82543955;
fax.: +86-10-82543977. Email address: wuzb@lnm.imech.ac.cn (Z.-B.
Wu).}

\newpage
\begin{abstract}
Thermocapillary migration of a planar non-deformable droplet in
flow fields with two uniform temperature gradients at moderate and
large Marangoni numbers is studied numerically by using the
front-tracking method. It is observed that the thermocapillary
motion of planar droplets in the uniform temperature gradients is
steady at moderate Marangoni numbers, but unsteady at large
Marangoni numbers. The instantaneous migration velocity at a fixed
migration distance decreases with increasing Marangoni numbers.
The simulation results of the thermocapillary droplet migration at
large Marangoni numbers are found in qualitative agreement with
those of experimental investigations. Moreover, the results
concerned with steady and unsteady migration processes are further
confirmed by comparing the variations of temperature fields inside
and outside the droplet. It is evident that at large Marangoni
numbers the weak transport of thermal energy from outside of the
droplet into inside cannot satisfy the condition of steady
migration process, which implies that the advection around the
droplet is a more significant mechanism for heat transfer
across/around the droplet at large Ma numbers. Furthermore, from
the condition of overall steady-state energy balance in the flow
domain, the thermal flux across its surface is studied for a
steady thermocapillary droplet migration in a flow field with
uniform temperature gradient. By using the asymptotic expansion
method, a nonconservative integral thermal flux across the surface
is identified in the steady thermocapillary droplet migration at
large Marangoni numbers. This nonconservative flux may well result
from the invalid assumption of quasi-steady state, which indicates
that the thermocapillary droplet migration at large Marangoni
numbers cannot reach steady state and is thus a unsteady process.


\end{abstract}

\newpage
\section{Introduction}
The transport phenomenon of droplets/bubbles in a liquid is a very
important topic for both fundamental hydrodynamics and practical
applications such as production of pure materials in manufacturing
industry and mass transfer in chemical engineering. Under normal
gravity, the motion of droplet/bubbles results from the buoyancy
when the densities of two fluids are different. With fast
development of space exploration, the studies on the physical
mechanism of droplet/bubble migration phenomena under reduced
gravity become more and more important. In this case, the buoyant
effect vanishes, the droplet/bubble moves as a result of the
variance of interface tension. Thus, in the microgravity
environment, a droplet/bubble suspended in an ambient fluid will
move in the direction of temperature gradient due to
thermocapillary force[1]. Thermocapillary motion of a single
droplet was firstly examined both theoretically and experimentally
by Young, Goldstein \& Block (1959)\cite{2}. They gave an
analytical prediction on its migration speed in the limit case of
zero Reynolds (Re) and zero Marangoni (Ma) numbers, which is
called as YGB model. Since then, the thermocapillary migration of
a bubble has been studied extensively by a series of theoretical
analyses\cite{3,4,5,6}, numerical simulations\cite{7,8,9,10} and
experimental investigations\cite{11}. In the mean time, several
numerical techniques for treating the two-phase flow, such as the
front-tracking method\cite{12,13} and the level-set
method\cite{14}, have also been developed, which may provide
effective techniques to directly investigate thermocapillary
migration processes of bubbles or droplets\cite{15,16,17,18},
interfacial mass transfer\cite{19,20} and interfacial flows with
soluble surfactants\cite{21,22}.

For the migration of a droplet, the experimental result for the
migration speed at small Re numbers obtained by Braun et
al(1993)\cite{23} agrees with the YGB model. To include small
inertial effects, the YGB model analysis was extended to the range
of small Ma numbers\cite{24}. For finite Ma numbers, several
numerical simulation on the three-dimensional thermocapillary
motion of non-deformable and deformable droplets were reported by
Wang et al\cite{25} and Haj-Hariri et al\cite{17}, respectively.
They used the front-tracking and level-set methods to catch the
interface and investigate the effects of physical parameters on
migration speeds and mobility, respectively. For large Ma numbers,
Balasubramanian \& Subramanian(2000)\cite{26} used thermal
boundary layers and found that the migration speed of a droplet
increased with increasing Ma number, which is in qualitative
agreement with corresponding numerical simulations\cite{27}. Both
the theoretical analysis and numerical simulation are based on the
assumptions of quasi-steady state and non-deformation of the
droplet. However, the experimental results of Hadland et
al(1999)\cite{11} and Xie et al(2005)\cite{28} were not in
qualitative agreement with the above theoretical and numerical
results, and it was shown that the droplet migration speed
non-dimensionalized by the YGB velocity decreased as Ma number
increased. The experiment investigation was completed in several
ranges of large Ma numbers, where the droplet migration was in an
accelerating state and did not reach a steady one. Recently, a
numerical investigation based on an axisymmetric droplet
model\cite{29,30} found that the steady thermocapillary migration
process did exist in a laboratory coordinate system, and verified
the above experimental results for the case of large Ma numbers,
however, the effects of Capillary numbers were not given in the
calculations. Owing to the invariance theory under transformation
between two inertia frames, the above numerical result in a
laboratory coordinate frame, i.e., the steady migration speed of
the droplet decreases as Ma increases, should be in agreement with
the theoretical and numerical results in a reference frame moving
with the steady droplet velocity\cite{26,27}. However, this seems
impossible. Moreover, Herrmann et al\cite{18} adopted numerical
method to investigate the thermocapillary motion of deformable
droplets and indicated that for large Ma numbers the assumption of
quasi-steady state was not valid. Therefore, the thermocapillary
droplet migration at large Ma numbers is still a topic to be
studied with emphasis laid on its physical mechanism.

The planar or cylindrical droplet/bubble as a simple model has
been extensively used to study its dynamical
mechanism\cite{31,32,33}. In this paper, we use the front-tracking
method to numerically study the thermocapillary migration of a
non-deformable planar droplet in the liquids at moderate and large
Ma numbers, and analyze in detail the relation of the migration
velocity to the temperature distribution inside and outside the
droplet. Moreover, by using asymptotic analysis, we investigate
the continuity of thermal flux across the surface based on the
overall energy balance of the droplet, and analyze the existence
of quasi-steady migration of the droplet at large Ma numbers.

\section{Governing equations}

Consider the thermocapillary migration of a planar droplet in a
continuous phase fluid of infinite extent under a uniform
temperature gradient $G$. Gravity and deformation of the droplet
shape are ignored. Two-dimensional continuous, momentum and energy
equations for the continuous phase fluid and the droplet in a
laboratory coordinate system are written as follows

\begin{equation}
\begin{array}{l}
\frac{\partial{\rho_i}}{\partial t} + \nabla \cdot (\rho_i {\bf
v}_i)
=0,\\
\frac{\partial\rho_i {\bf v}_i}{\partial t} + \nabla \cdot (\rho_i
{\bf v}_i{\bf v}_i) = - \nabla p_i + \nabla \cdot \mu_i(\nabla
{\bf v}_i +\nabla^T
{\bf v}_i + {\bf F}_{\sigma},\\
\frac{\partial{T_i}}{\partial t} + \nabla \cdot ({\bf v}_i T_i)
=\frac{\kappa_i}{k_i} \nabla \cdot ({k_i \nabla T_i}),
\end{array}
\end{equation}
where ${\bf v}_i$ and $T_i$ are velocity and temperature,
respectively. ${\bf F}_{\sigma}$ is the surface tension force
acting on the interface.
 $\rho_i$, $\mu_i$, $k_i$,
$\kappa_i$ represent density, dynamic viscosity, thermal
conductivity, and thermal diffusivity, respectively. Symbols with
subscript 1 and 2 denote physical coefficients of the continuous
fluid and the droplet, respectively. The solutions of Eqs. (1)
have to satisfy the boundary conditions at infinity
\begin{equation}
{\bf v}_1=0, T_1 \to  T_0 + Gz,
\end{equation}
where $T_0$ is the undisturbed temperature of the continuous phase
and the boundary conditions at the interface $(r_b, z_b)$ of the
two fluids
\begin{equation}
\begin{array}{l}
{\bf v}_1(r_b, z_b,t)={\bf v}_2(r_b, z_b,t),\\
T_1(r_b, z_b,t) =T_2(r_b,z_b,t),\\
k_1 \frac{\partial{T_1}}{\partial n}(r_b, z_b,t) = k_2
\frac{\partial{T_2}}{\partial n}(r_b, z_b,t),
\end{array}
\end{equation}
where ${\bf n}$ is a unit vector normal to the interface. In the
modelling assumptions, both fluids are immiscible and the physical
properties are constant. The equations of state for density,
viscosity, heat conduction and heat diffusivity are written as
follows

\begin{equation}
\frac{d \rho_i}{d t} =\frac{d \mu_i}{d t} =\frac{d k_i}{d t}
=\frac{d \kappa_i}{d t} =0.
\end{equation}
The reference velocity is defined as
\begin{equation}
v_o=-\sigma_T G R_0/\mu_1,
\end{equation}
where $R_0$ is the radius of the droplet, and
$\sigma_T(=d\sigma/dT)$ is the change rate of interfacial tension
with temperature.
 By taking $R_0$, $v_o$ and $GR_0$ as the characteristic quantities to make coordinates,
velocity and temperature dimensionless, Eqs. (1) are rewritten in
the following non-dimensional form
\begin{equation}
\begin{array}{l}
 \nabla \cdot {\bf v}_i=0,\\
\frac{\partial\rho_i {\bf v}_i}{\partial t} + \nabla \cdot (\rho_i
{\bf v}_i{\bf v}_i) = - \nabla p_i + \frac{1}{Re} \nabla \cdot
\mu_i(\nabla
{\bf v}_i +\nabla^T {\bf v}_i) + {\bf f}_{\sigma},\\
\frac{\partial T_i}{\partial t} + \nabla \cdot ({\bf v}_i T_i) =
\frac{1}{Ma} \frac{\kappa_i}{k_i} \nabla \cdot ({k_i \nabla T_i}),
\end{array}
\end{equation}
where the physical coefficients are non-dimensionlized by the
characteristic quantities of continuous fluid  and ${\bf
f}_{\sigma}={\bf F}_{\sigma}R_0/\rho_1 v_0^2$.  The boundary
conditions (2) and (3) are non-dimensionlized as
\begin{equation}
\begin{array}{l}
{\bf v}_1=0, T_1 \to T_0 +  z,
\end{array}
\end{equation}
at infinity and
\begin{equation}
\begin{array}{l}
{\bf v}_1(r_b, z_b,t)={\bf v}_2(r_b, z_b,t),\\
T_1(r_b, z_b,t) =T_2(r_b, z_b,t),\\
k_1 \frac{\partial{T_1}}{\partial n}(r_b, z_b,t) = k_2
\frac{\partial{T_2}}{\partial n}(r_b, z_b,t),
\end{array}
\end{equation}
at the interface between two fluids.  The Reynolds number and
Marangoni number are defined as
\begin{equation}
 Re=\frac{\rho_1 v_0 R_0}{\mu_1},\ \ \  Ma=\frac{v_0R_0}{\kappa_1} = Pr Re,
\end{equation}
where $Pr=\mu_1/\rho_1 \kappa_1$ is the Prandtl (Pr) number. In
what follows, the undisturbed temperature $T_0$ and
non-dimensional physical parameters ($\rho_1= \mu_1= k_1=
\kappa_1=1$) of the continuous phase are reduced for simplicity,
except when otherwise indicated.

\section{Numerical simulation of thermocapillary droplet migration at moderate and large Ma numbers}

\subsection{Models and numerical methods}

As shown schematically in Fig.1, the symmetric axis of the
container is taken as the $z$-axis. A droplet is placed initially
at the center of coordinates and then moved along the $z$-axis.
Thus, the solution of Eqs. (6) should satisfy the following
initial conditions in the whole domain $x \in [x_0,x_1]$ and $z
\in [z_0,z_1]$
\begin{equation}
{\bf v}_i =0, \ \ \ T_i= z
\end{equation}
and  non-slip/periodic boundary conditions at the top and bottom
walls/the horizontal boundaries
\begin{equation}
\begin{array}{l}
{\bf v}_1(x,z_0) ={\bf v}_1(x,z_1) =0, \ \ \ T_1(x,z_0) =z_0, T_1(x,z_1) =z_1,\\
{\bf v}_1(x_0,z) ={\bf v}_1(x_1,z), \ \ \ T_1(x_0,z) = T_1(x_1,z).
\end{array}
\end{equation}

 In the computation, we use a
fixed regular staggered MAC grid in the computational domain. To
discretize Eqs. (6), we adopt a second-order central difference
scheme for the spatial variables and an explicit
predictor-corrector second-order scheme for time integration. The
predictor-corrector method is a combination of the explicit Euler
and the implicit trapezoidal methods to obtain an explicit
technique with better convergence characteristic. In the method,
the solution at time step $n+1$ is predicted by using the explicit
Euler method
\begin{equation}
{\phi}^\dag_{n+1} = \phi^n + f(t_n,\phi^n) \Delta t,
\end{equation}
where $\dag$ indicates that this is not the final value of the
solution at $t_{n+1}$. Rather, the solution is corrected by
applying the trapezoid rule
\begin{equation}
\phi^{n+1} = \phi^n + \frac{1}{2} [f(t_n,\phi) +
f(t_{n+1},\phi^\dag_{n+1})] \Delta t.
\end{equation}
To achieve the second-order accurate time integration of the
velocity and temperature fields in Eqs. (6), we employ the
Chorin's projection method to outline the first-order Euler
integration in (12) as follows.

Since both fluids are assumed immiscible, all physical
coefficients are discontinuous across the interface. The interface
is captured and updated by the front-tracking method\cite{12,13}.
 When the interface is moved to
a new position, the density field is updated. The interface is
considered to have a finite width so that the density across the
interface is continuous. Here, a weighting function suggested by
Peskin\cite{34} is adopted as
 \begin{equation}
 w_{ij}({\bf r}_p) =d(x_p-i\Delta x)d(z_p-j\Delta z),
 \end{equation}
where

\begin{equation}
 d(r) = \left \{ \begin{array}{ll}
 (1/4 \Delta r)[1 +\cos (\pi r/2h)], &|r|<2 \Delta r,\\
  0,                                 &|r| \ge 2 \Delta r,
                 \end{array} \right.
\end{equation}
and $(x_p, z_p)$ is the interface node. Once the density is
updated, the velocity field will be computed by the Chorin's
projection method, which is divided into two steps. One is a
prediction step, where the effect of the pressure is ignored
\begin{equation}
\frac{\rho_i^{n+1}{\bf v}_i^* -\rho_i^n{\bf v}_i^n}{\Delta t} =
-\nabla \cdot (\rho_i^n {\bf v}_i^n{\bf v}_i^n)  + \frac{1}{Re}
\nabla \cdot \mu_i^n (\nabla {\bf v}_i^n + \nabla^T {\bf v}_i^{n})
+{\bf f}_\sigma.
\end{equation}
Another is a correction step in terms of the pressure gradient
\begin{equation}
\frac{\rho_i^{n+1} {\bf v}_i^{n+1} -\rho_i^{n+1}{\bf
v}_i^*}{\Delta t} =-\nabla p_i^{n+1},
\end{equation}
where the pressure is obtained by solving the following Poisson
equation
\begin{equation}
\nabla \cdot \frac{1}{\rho_i^{n+1}} \nabla p_i^{n+1} =
\frac{1}{\Delta t} \nabla \cdot {\bf v}_i^*.
\end{equation}
In solving Eq. (18), we use the successive over relaxation
iteration method to get $p_i^{n+1}$. When the pressure is
obtained, the corrected velocity field ${\bf v}_i^{n+1}$ is
determined from Eq. (17). Similarly, the energy equation is
discretized in the form
\begin{equation}
\frac{T_i^{n+1} -T_i^n}{\Delta t} = -\nabla \cdot ({\bf v}_i^{n+1}
T_i^n) + \frac{1}{Ma}\frac{\kappa_i^n}{k_i^n} \nabla \cdot (k_i^n
\nabla T_i^n).
\end{equation}
In terms of the corrected velocity field ${\bf v}_i^{n+1}$, the
temperature field $T_i^{n+1}$ is determined. Until now, by using
the projection method, the first-order time integration
${\phi}^\dag_{n+1}$ of the velocity, pressure and temperature
fields is completed. In the mean time, other physical coefficients
$(\mu, k, \kappa)$ across the interface at the time step $n+1$ are
also updated to have the same distribution as the density.
Repeating the above process, we get a second first-order accurate
solution ${\phi}^{\dag\dag}_{n+2}$ at time step $n+2$ based on the
first-order accurate solution ${\phi}^\dag_{n+1}$
\begin{equation}
{\phi}^{\dag\dag}_{n+2} = \phi^\dag_{n+1} +
f(t_{n+1},\phi^\dag_{n+1}) \Delta t.
\end{equation}
Finally, the solution for the second-order time integration is
obtained as follows
\begin{equation}
{\phi}^{n+1} = (\phi^n + \phi^{\dag\dag}_{n+2})/2.
\end{equation}
 Since the droplet in the migration process is assumed
 non-deformable, the vertical area average velocity in the droplet is taken as
 the droplet migration velocity $V_z$. The nodes of interface are moved
 with this velocity at each time step.

In solving Eq. (16), the surface tension force ${\bf f}_\sigma$ is
determined by referring to the temperature field. In general, the
surface tension force on a short front element is defined as
 \begin{equation}
 \delta {\bf F}_{\sigma} = \int_{\Delta s} \frac{\partial}{\partial s} (\sigma {\bf
 t})ds =(\sigma {\bf t})_2 -(\sigma {\bf t})_1 =\Delta_{21}(\sigma {\bf t}),
 \end{equation}
where ${\bf t}$ is an unit tangent vector, $s$ is the arc length
along the interface. $\sigma$ is the  surface tension coefficient
written as
\begin{equation}
\sigma = \sigma_0 + \sigma_T T,
\end{equation}
where $\sigma_0$ is the surface tension coefficient at a reference
temperature $T_0$, and $\sigma_T$ is a negative constant for most
fluids. By using the above characteristic quantities, the
non-dimensional surface tension force is written in the form of
body force as
\begin{equation}
\begin{array}{ll}
 \delta {\bf f}_{\sigma} &= \delta {\bf F}_{\sigma} (R_0/ \rho_1 v^2_0)/(R^2_0 \delta x \delta
 z)\\
 &= \Delta_{21} (\sigma {\bf t}) /\rho_1 v^2_0 R_0 \delta x \delta z\\
 &= \Delta_{21} [({\sigma_0/v_0 \mu_1 - T}){\bf t}] /Re \delta x \delta z\\
 &= \Delta_{21} [(1/Ca -T) {\bf t} ]/Re \delta x \delta z,
 \end{array}
 \end{equation}
where Ca(=$v_0 \mu_1/\sigma_0$) is the Capillary number. To
calculate the surface tension force ${\bf f}_\sigma$,
 the surface temperature on the interface is firstly obtained
 by interpolating values on the grid points. The tangent vector
 is computed from a Lagrangian polynomial fitting through four interface nodes. Then,
 the surface tension force on the interface is distributed to
 the grid points by means of weighting function (14).

\subsection{Results and analysis}

To check the sensitivity of the results to grid refinements,  we
perform calculations for a planar droplet migration at Re=5,
Ma=20, Ca=0.01666 and $\rho_2$= $\mu_2$= $k_2$= $\kappa_2$=0.5
using the method described above. The computational domain is
chosen as $4 \times 8$. Based on $64 \times 128$, $96 \times 192$
and $128 \times 256$ grid points, i.e., 16, 24 and 32 grid points
per droplet radius, the time evolution of the droplet migration
velocity is calculated and plotted in Fig. 2. The migration
velocity curve seems to converge when the grid becomes finer. The
difference in the migration velocities computed with 24 and 32
grid points per droplet radius is very small (about $1.5\%$). In
the following calculations we fix 24 grid points per droplet
radius as the grid resolution. To further validate our code, we
compare the current computation results with Nas \&
Tryggvason's\cite{15}, where the deformation of the droplet is
considered. In Fig.~3, it is observed that both results have the
same trends and the migration velocities are close together.

In the following calculations, we adopt the silicone oil of
nominal viscosity 5cst and the FC-75 Fluorinert liquid, i.e., the
working media in the space experiment\cite{28}, as the continuous
phase fluid and the droplet, respectively. The physical parameters
of the continuous fluid and the droplet at temperature $25^o$C are
given in Table I. $\sigma_T$ is fixed as -0.044 dyn/cmK\cite{28}
and $\sigma_0 \approx$ 6 dyn/cm\cite{35} is adopted. From the
values of the continuous fluid parameters, the Pr number and the
capillary length $\lambda_0=\sqrt{\sigma_0/\rho_1g_0}$ (with the
earth's gravity $g_0=980 cm/s^2$) are determined as 67.8 and
0.08cm, respectively. The most of droplets in the space
experiment\cite{28} have $R_0 \geq \lambda_0$, which refers to the
domination of the gravitational effect to the droplet shapes on
the earth. However, in the microgravity environment(the effective
gravity $g_e$ is about $O(10^{-6})$ of $g_0$), the gravitational
effect is neglected($R_0 \ll \lambda_e$), so the droplet shapes
are dominated by the capillary effect. The computational domain is
chosen as $\{x,z\} \in \{[-4,4], [-4,16]\}$ and the resolution is
fixed at $192 \times 480$. The initial droplet is placed at the
position (0,0) and the time step is 0.0002.

\subsubsection{Flow field with the temperature gradient $G$=12 K/cm}

In the space experiments with $G$=12 K/cm\cite{28}, Re and Ma lie
respectively in the ranges of 4.5-302.6 and 145-5525, their
specific values depending on $R_0$. To simulate the experimental
processes,
 the physical coefficients in the droplet migration processes
are determined by changing $R_0$. The correspondence of Re, Ma and
Ca to $R_0$ is presented in Table II, where Re is in a range of
moderate values and Ma have both moderate and large numbers. Fig.
4 displays the time evolution of droplet migration velocities for
five sets of non-dimensional coefficients. In the present range of
Ma, the migration velocities versus time have complex behaviors,
which can be classified into three types based on the curve
characters. At Ma=44.7(Re=0.66), the initial migration velocity
increases sharply before $t=3$, and then drops to approach a
steady value. For Ma=402.5-1118.1(Re=5.93-16.5), the initial
accelerating process has smaller peak value as Ma increases. After
the increasing-decreasing oscillation process,
 the terminal droplet migration velocity increases with time,
 i.e., the droplet migration is in an accelerating state.
 The slope of the curve increases as Ma increases.
 For Ma=2191.6-3622.8(Re=32.3-53.4), the droplet migration
velocity increases monotonously with time and decreases with
increasing Ma. We can thus conclude that in the time frame under
investigation the thermocapillary droplet migration is steady at
moderate Ma numbers, but becomes unsteady at large Ma numbers. In
the two space experiments, Figs. 4 of \cite{11} and \cite{28}
showed that the whole migration processes were unsteady and didn't
reach any steady state. Even a plateau appears in the curve of
migration velocity vs migration distance, the migration process
seems to be an accelerating one after the slow varying period. To
further compare with the experimental investigation\cite{28}, we
take several fixed migration distances $l_z$ and determine the
relation between instantaneous non-dimensional migration velocity
$V_z$ and Ma numbers. The numerical and experimental results are
plotted in Fig. 5, from which it is evident that both the
numerical and experimental migration velocities of droplet
decrease as Ma increases in the range of large Ma numbers. Hence,
at large Ma numbers, the above simulation results are in
qualitative agreement with those of experimental
investigations\cite{11,28}.

In order to understand the phenomena exhibited in droplet
migration processes, it is important to analyze the evolution of
the velocity and temperature fields. Fig. 6 displays the computed
velocity fields at $t=20$ in both the laboratory coordinate frame
and the reference frame moving with the droplet at
Re=16.5(Ma=1118.1). In the laboratory coordinate frame, the
streamlines for a moving droplet are closed and symmetric about
the $z$-axis. In the reference frame, when the external
streamlines go around the droplet, a pair of vortices is formed
inside the droplet. It is evident that in the reference frame
recirculation flows in both the continuous phase fluid and the
droplet are driven by the surface tension force generated by the
temperature gradient along the surface. In Fig. 7, we depict the
pattern evolution of streamlines with time in a reference frame
moving with the droplet at Re=16.5(Ma=1118.1). Initially, there
appear two vortices symmetric about the vertical diameter. Along
with the rising of the droplet, the pair of vortices in the
droplet is kept, but the vortex centers are moving up. In the
whole process, the basic types of streamlines for both the
internal and external motions are not changed. For moderate Re
numbers, both the convection and viscous terms in the momentum
equation have important effects on the fluid flow. The external
flow just passes around the droplet and does not separate from the
droplet surface, and thus the computed velocity fields are similar
in the range of moderate Re numbers.

In the range of $R_0$, the fluid flow has behaviors similar to
those for moderate Re numbers, but the thermal transfer exhibits
different characters for moderate and large Ma numbers. For
moderate Ma numbers, both the heat convection and the heat
conduction have important effects on the energy transfer. Fig. 8
displays the time evolution of isotherms at moderate Ma (Ma=44.7,
Re=0.66), which corresponds to that of migration velocities shown
in Fig. 4. As given in Eq. (23), the surface tension coefficient
decreases with the increasing of the local temperature. For a
temperature field with its gradient in the $z$ direction, the
generated surface tension force is a net force along the surface.
At the beginning, the droplet starts to move towards the warm side
under the action of net force. It induces in turn viscous stresses
in both fluids, which causes streamlines inside and outside the
droplet to form double vortices and to go around the droplet,
respectively. The temperature field inside the droplet is affected
by the two rotating vortices. The horizontal isotherm $T=0$ in the
droplet is moving up, as well as bending along the migration
direction. Along with the rising of the droplet, both the internal
and external temperature fields around the droplet surface are
re-distributed due to the action of heat convection. In the
process, the isotherm $T=0$ moves up and approaches the top of the
droplet. At $t=20$, a small cap-type isotherm $T=0$ is formed
within the droplet. Meanwhile, under the action of heat
conduction, the thermal energy is transferred from outside of the
droplet to inside, so the temperature inside the droplet
increases. At $t=60$, the temperature of cap-type isotherm in the
droplet reaches $T=2$. Hence, both the temperature fields inside
and outside the droplet increase with time. Only when both the
internal and external temperature fields satisfy linear relations
with the same slope, does the droplet migration reach a steady
state. For large Ma numbers, the effect of heat convection is
stronger than that of heat conduction. Fig. 9 displays the time
evolution of isotherms at large Ma (Ma=402.5, Re=5.93), which
corresponds to that of migration velocity shown in Fig. 4. In the
initial (accelerating) stage of migration, the evolution of the
isotherms is similar to that shown in Fig. 8. In the following
(oscillation) stage of migration, the bent isotherms in the
droplet move to approach the top of the droplet and are converted
to pea-type ones, and then to earphone-type ones. In the last
(accelerating) stage of migration, the earphone-type isotherms are
transformed into those with two symmetric vortices. In the whole
migration process, although the temperature inside the droplet
increases, the temperature of minimal isotherm is still kept at
$T=0$. It implies that the thermal energy transfer from outside of
the droplet to inside is weaker than that shown in Fig. 8. Fig. 10
displays the time evolution of isotherms at a still larger Ma
(Ma=2191.6, Re=32.3), which corresponds to that of migration
velocity shown in Fig. 4. In the whole (accelerating) process of
migration, the initial evolution of the isotherms is similar to
that shown in Fig. 9, except for the slower movement of the
isotherms in approaching the top of the droplet. Then, the
pea-type isotherms are converted to earphone-type ones. And
finally, the isotherms with two vortices symmetric about the
vertical diameter in the droplet are formed. At $t=60$, the fact
that the minimal isotherm $T=0$ has larger closed area in the
droplet means that the thermal energy transfer from outside of the
droplet to inside is weaker than that shown in Fig. 9. Thus, at
large Ma numbers, although the temperature outside the droplet
increases fast as the droplet rises, the temperature inside the
droplet has only a slow increase. The droplet migration does not
reach a steady state, and is thus a unsteady process.

To further quantitatively depict the steady and unsteady migration
processes, we will investigate the time evolution of temperature
fields inside and outside the droplet. Fig. 11 displays the
temperature at the point $(x_c,z_c)$ inside the droplet and the
point $(x_c,z_c+2)$ outside the droplet
 in the migration processes, where
$(x_c,z_c)$ is the center of the droplet in the laboratory
coordinate system. It is evident that for the moderate Ma(=44.9)
number time evolution curves of the temperature at these two
points after $t_0=20$ are approximately linear and parallel.  Both
the temperature inside and outside the droplet satisfy the linear
relation: $T_i=T_i(t_0) + V_\infty (t-t_0)$, which indicates a
steady migration process with the constant velocity
$V_z=V_\infty$. However, for large Ma numbers, although the time
evolution curves of the temperature inside and outside the droplet
after $t_0=20$ are approximately linear, but they are not
parallel. The slope of the time evolution curve of the temperature
inside the droplet is smaller than that outside the droplet, so
the difference of the temperatures at these two points increases
as time increases. It implies that the terminal droplet migration
does not reach a steady state, and is thus a unsteady process.
Therefore, the advection around the droplet is a more significant
mechanism for heat transfer across/around the droplet at large Ma
numbers.

\subsubsection{Flow field with the temperature gradient $G$=9 K/cm}

In the space experiments with $G$=9 K/cm\cite{28}, Re and Ma lie
respectively in the ranges of 3.2-89.8 and 148-4103, their
specific values depending on $R_0$. To simulate the experimental
processes,
 the physical coefficients in the droplet migration processes
are determined by changing $R_0$. The correspondence of Re, Ma and
Ca to $R_0$ is presented in Table III, where Re is in a range of
moderate values and Ma have both moderate and large numbers. Fig.
12 displays the time evolution of droplet migration velocities for
five sets of non-dimensional coefficients. In the present range of
Ma, the curves of migration velocities versus time are classified
into three types based on their characters.  At Ma=33.5(Re=0.49),
the initial migration velocity increases sharply near $t=2$, and
then drops to approach a steady value. For
Ma=301.9-838.6(Re=4.45-12.4), after the increasing-decreasing
oscillation process, the droplet migration is in an accelerating
state. For Ma=1643.6-2717.1(Re=24.2-40.1), the droplet migration
velocity increases monotonously with time and decreases with
increasing Ma. We can thus conclude that in the time frame under
investigation the thermocapillary droplet migration is steady at
moderate Ma numbers, but becomes unsteady at large Ma numbers. To
further compare with the experimental investigation\cite{28}, we
take several fixed migration distances $l_z$ and determine the
relation between instantaneous non-dimensional migration velocity
$V_z$ and Ma numbers. The numerical and experimental results are
plotted in Fig. 13, from which it is evident that both the
numerical and experimental migration velocities of droplet
decrease as Ma increases in the range of large Ma numbers. Hence,
at large Ma numbers, the above simulation results are in
qualitative agreement with those of experimental investigations.

\section{Theoretical analysis of thermocapillary droplet migration at large Ma numbers}

\subsection{Quasi-steady state assumption}
In general, the surface tension is a linear decreasing function of
the local temperature. For a temperature field with its gradient
in the $z$ direction, the generated surface tension force is a net
force along the surface and the droplet starts to move towards the
warm side under the action of net force. When the net force acting
on the droplet at the flow direction is zero, the thermocapillary
droplet migration reaches a steady process. However, due to the
variation of physical parameters with the ambient temperature, the
migration process may not reach any steady state. Only when
 the migration is sufficiently slow that the order of
 relevant time scale for the transport process to generate
 steady velocity and temperature fields is
 smaller than that for the droplet to move an appreciable distance,
 the assumption of the quasi-steady state is valid.
It means that after experiencing an initial unstable migration
process, the droplet migration may reach a steady state at the
time $t_0$ and the position ${\bf r}_0=z_0 {\bf k}$, i.e.,
migrating with a constant droplet migration speed $V_{\infty}$.
Using the coordinate transformation from the laboratory coordinate
system to a coordinate system moving with the droplet velocity
$V_{\infty}$
\begin{equation}
\begin{array}{lll}
{\bf r} = \bar{\bf r}  + {\bf r}_0 +V_{\infty}(t-t_0) {\bf k},&
{\bf v}_i({\bf r},t) = \bar{\bf v}_i(\bar{\bf r})  + V_{\infty}
{\bf k},& T_i({\bf r},t) = \bar{T}_i(\bar{\bf r})  + z_0 +
V_{\infty}(t-t_0),
\end{array}
\end{equation}
 the energy equation in Eqs.(6) can be formulated as
\begin{equation}
\begin{array}{l}
V_{\infty} + {\bar \nabla} \cdot (\bar{\bf v}_i \bar{T}_i)=
\frac{1}{Ma} \frac{\kappa_i}{k_i} \bar{\nabla} \cdot (k_i
\bar{\nabla} \bar{T}_i),
\end{array}
\end{equation}
in a polar coordinate system $\bar{\bf r}=(\bar{r}, \theta)$.
Under the assumption of non-deformable droplet, the radial
coordinate axis is the outer normal vector of the interface. By
using the transformation (25), the boundary conditions (7) and (8)
can be respectively written as follows
\begin{equation}
\bar{\bf v}_1=-V_{\infty}{\bf k}, \bar{T}_1 \to \bar{r} \cos
\theta,
\end{equation}
 at places far away from the droplet and
\begin{equation}
\begin{array}{l}
\bar{u}_{1n}(1,\theta)=\bar{u}_{2n}(1,\theta)=0,\\
\bar{v}_{1s}(1,\theta)=\bar{v}_{2s}(1,\theta),\\
\bar{T}_1(1,\theta) =\bar{T}_2(1,\theta),\\
\frac{\partial{\bar{T}_1}}{\partial n}(1,\theta) = k_2
\frac{\partial{\bar{T}_2}}{\partial n}(1,\theta)
\end{array}
\end{equation}
at the interface of the two fluids. Thus, once the droplet
migration reaches a steady state, the above problem of (6)(7)(8)
in the laboratory coordinate system can be described by steady
energy equations (26) with boundary conditions (27)(28) in the
coordinate system moving with the droplet velocity. This implies
the overall steady-state energy balance with two phases in the
flow domain in the co-moving frame of reference.

\subsection{Nonconservative integral thermal flux across the droplet
surface at large Ma numbers}
 To analyze the energy equations with
a small parameter $\epsilon=1/\sqrt{V_\infty Ma}$, Eqs. (26) are
rewritten as
\begin{eqnarray}
1+\bar{\nabla} \cdot (\bar{\bf v}_1 \bar{T}_1) =
\epsilon^2 \bar{\nabla} \cdot (\bar{\nabla} \bar{T}_1),\\
1+\bar{\nabla} \cdot (\bar{\bf v}_2 \bar{T}_2) = \epsilon^2
\kappa_2 \bar{\nabla} \cdot (\bar{\nabla} \bar{T}_2),
\end{eqnarray}
where both the velocity fields $\bar{\bf v}_i$ in the surrounding
fluids and in the droplet are rescaled by $V_\infty$.

To confirm the overall steady-state energy balance of two phases
in the flow domain with respect to the co-moving frame of
reference, we have to integrate Eqs. (29)(30) with an asymptotic
expansion of the outer temperature field at infinity with respect
to the small parameter $\epsilon$. To determine the asymptotic
behavior of $\bar{T}_1$ at $\bar{r} \gg 1$, we rewrite Eq. (29) as
follows
\begin{equation}
1+ \bar{u}_1 \frac{\partial \bar{T}_1}{\partial \bar{r}}
+\frac{\bar{v}_1}{\bar{r}}
\frac{\partial{\bar{T}_1}}{\partial{\theta}}= \epsilon^2
 \bar{\Delta} \bar{T}_1.
\end{equation}
Let
\begin{equation}
\begin{array}{l}
\bar{u}_1 = \bar{u}^0_1 + o(1),\\
\bar{v}_1 = \bar{v}^0_1 + o(1),\\
\bar{T}_1 = \bar{T}^0_1 + o(1),
\end{array}
\end{equation}
we have the energy equation to leading order
\begin{equation}
1+ \bar{u}^0_1 \frac{\partial \bar{T}^0_1}{\partial \bar{r}}
+\frac{\bar{v}^0_1}{\bar{r}}
\frac{\partial{\bar{T}^0_1}}{\partial{\theta}}=0.
\end{equation}
By using the characteristic line method, the primary approximation
of the outer temperature field
 in the continuous phase is
 derived as
\begin{equation}
\bar{T}_1=\bar{r} \cos \theta + \int^{\bar{r}}_\infty (\bar{v}^0_1
\sin \theta -\bar{u}^0_1 \cos \theta -1)/\bar{u}^0_1|_\Psi
d\tilde{r} + o(1),
\end{equation}
where $\Psi \sim (\bar{r}-1/\bar{r})\sin \theta$. For moderate Re
numbers, the velocity fields in Eq. (34) can be described by the
potential flows. By using the scaled inviscid velocity field in
the continuous phase flow passing a circular cylinder\cite{MT}
\begin{equation}
\begin{array}{l}
\bar{u}^0_1=-\cos \theta (1-\frac{1}{\bar{r}^2}),\\
\bar{v}^0_1=\sin \theta (1+\frac{1}{\bar{r}^2}),
\end{array}
\end{equation}
Eq. (34) is written as
\begin{equation}
\bar{T}_1=\bar{r} \cos \theta + \int_{\bar{r}}^\infty
\frac{1}{\tilde{r}^2-1} \frac{2 \Psi^2 /
(\tilde{r}-1/\tilde{r})^2-1}{\pm
\sqrt{1-\Psi^2/(\tilde{r}-1/\tilde{r})^2}}|_\Psi d\tilde{r} +
o(1),
\end{equation}
where $\Psi[=\sin \theta(\bar{r}-1/\bar{r})]$ is streamfunction of
the continuous phase, the symbol "+" before the integral is so
determined as to preserve the monotonously increasing trend of
$\bar{T}_1(\bar{r},0)$ with $\bar{r}(>1)$ in the continuous phase
and the symbol "$\pm$" in the integral depends on the value of
$\theta$ (the symbol $"+"/"-"$ corresponds to $\theta \in
[0,\pi/2)/[\pi/2,\pi)$). At $\bar{r} \gg 1$, Eq. (36) can be
expressed as
\begin{equation}
\begin{array}{ll}
\bar{T}_1 &\approx \bar{r} \cos \theta + \int_{\bar{r}}^\infty
\frac{1}{\tilde{r}^2} \frac{2 \Psi^2 / \tilde{r}^2-1} {\pm
\sqrt{1-\Psi^2/\tilde{r}^2}}|_\Psi
d\tilde{r} + o(1)\\
 &= \bar{r} \cos \theta -\frac{1}{\bar{r}} \cos \theta + o(1),
 \end{array}
\end{equation}
where $\Psi \approx \sin \theta \bar{r}$.

In physics, the thermocapillary migration of a planar droplet has
the mirror symmetry about the coordinate axis $\theta=0$ or $\pi$,
so the overall energy balance in the whole flow domain ${\theta
\in [0,2\pi)}$ can be generated by combining the energy balance in
two connected flow domains $\theta \in [0,\pi)$ and $\hat{\theta}
\in [0,\pi)$ through the transformation $\hat{\theta}=2\pi -
\theta$.
 Integrating Eq. (29) and Eq. (30) in the continuous phase
domain $(\bar{r}\in [1,\bar{r}_{\infty}],\theta\in[0,2\pi])$
  and within
the droplet region $(\bar{r}\in [0,1],\theta\in[0,2\pi])$, and
then transforming them to linear integration on the droplet
surface and the surface at infinity by using the Green's formula,
we have
\begin{equation}
\pi (\bar{r}^2_\infty-1) + \oint \bar{u}_{1n}
\bar{T}_1|_{\bar{r}_{\infty}} ds - \oint \bar{u}_{1n} \bar{T}_1|_1
ds = \epsilon^2 (\oint \frac{\partial \bar{T}_1}{\partial
n}|_{\bar{r}_{\infty}} ds - \oint
\frac{\partial \bar{T}_1}{\partial n}|_1 ds)\\
\end{equation}
and
\begin{equation}
\pi + \oint \bar{u}_{2n} \bar{T}_2|_1 ds= \epsilon^2 \kappa_2
\oint \frac{\partial \bar{T}_2}{\partial n}|_1 ds.
\end{equation}
Using the normal velocity boundary condition at the interface in
(28) and the temperature field at the infinity in (37), we can
derive

\begin{equation}
\oint \frac{\partial{\bar{T}_1}}{\partial{n}}|_1 ds =
-\frac{\pi}{\epsilon^2} (1-\frac{1}{\bar{r}^2_{\infty}}) +
o(\frac{1}{\epsilon^2})
\approx -\frac{\pi}{\epsilon^2}\\
\end{equation}
and
\begin{equation}
\oint \frac{\partial{\bar{T}_2}}{\partial{n}}|_1 ds =
\frac{\pi}{\kappa_2 \epsilon^2}.
\end{equation}
 To analyze
the thermal flux near the boundary, we write the integrals of Eq.
(40) and Eq. (41) in their discretization forms and simplify the
expressions in terms of the mirror symmetrical relationship
$\frac{\partial{\bar{T}_i}}{\partial{\bar{r}}}|_{1,\theta}
=\frac{\partial{\bar{T}_i}}{\partial{\bar{r}}}|_{1,2\pi-\theta}$
as
\begin{equation}
\oint \frac{\partial{\bar{T}_1}}{\partial{n}}|_1 ds =\int_0^{2\pi}
\frac{\partial{\bar{T}_1}}{\partial{\bar{r}}}|_1 d \theta =
\sum_{i=1}^N \frac{\partial{\bar{T}_1}}{\partial{\bar{r}}}|_1
\Delta \theta =2\sum_{i=1}^{N/2}
\frac{\partial{\bar{T}_1}}{\partial{\bar{r}}}|_1 \Delta \theta
 <0\\
\end{equation}
and
\begin{equation}
\oint \frac{\partial{\bar{T}_2}}{\partial{n}}|_1 ds =\int_0^{2\pi}
\frac{\partial{\bar{T}_2}}{\partial{\bar{r}}}|_1 d \theta =
\sum_{i=1}^N \frac{\partial{\bar{T}_2}}{\partial{\bar{r}}}|_1
\Delta \theta =2\sum_{i=1}^{N/2}
\frac{\partial{\bar{T}_2}}{\partial{\bar{r}}}|_1 \Delta \theta
>0,
\end{equation}
where  $\Delta \theta =2\pi/N$. And thus we arrive at a conclusion
that there must be some interface points $\theta_i \in [0,\pi]$
where the following equation holds
\begin{equation}
\frac{\partial{\bar{T}_1}}{\partial{\bar{r}}}(1,\theta_i)  < 0  <
 \frac{\partial{\bar{T}_2}}{\partial{\bar{r}}}(1,\theta_i)
\end{equation}
or some interface points $\theta_i$ and $\theta_j \in [0,\pi]$
where the following equations hold
\begin{equation}
\begin{array}{l}
0< \frac{\partial{\bar{T}_1}}{\partial{\bar{r}}}(1,\theta_i) <
 \frac{\partial{\bar{T}_2}}{\partial{\bar{r}}}(1,\theta_i),\\
\frac{\partial{\bar{T}_1}}{\partial{\bar{r}}}(1,\theta_j)   <
 \frac{\partial{\bar{T}_2}}{\partial{\bar{r}}}(1,\theta_j)<0.
 \end{array}
\end{equation}
 Physically, this means that near these points $\theta_i$ the thermal
energy is transferred from the interface to outside (the
surrounding fluid) as well as from the interface to inside (the
droplet) or near these points $\theta_i$/$\theta_j$ the
transference of thermal energy from outside/the interface to the
interface/outside is weaker/stronger than that from the
interface/inside to inside/the interface. On the one hand, if Eq.
(44) can satisfy the thermal flux boundary condition in Eqs. (28),
thermal sources inside the interface will be introduced to balance
the transference of thermal energy. On the other hand, if Eqs.
(45) can satisfy the thermal flux boundary condition in Eqs. (28),
thermal sinks inside the interface or thermal sources in the
droplet will be introduced to decrease the transference of thermal
energy from the interface to outside or increase the transference
of thermal energy from inside to the interface. Since there is
absolutely no thermal sources or sinks inside the interface or
thermal sources in the droplet, the above transport processes of
thermal energy near the interface seem impossible. It means that
the thermal flux across the droplet surface is nonconservative.
Moreover, from Eq. (40) and Eq. (41), we have
\begin{equation}
\oint [k_2 \frac{\partial{\bar{T}_2}}{\partial{\bar{r}}}|_1 -
\frac{\partial{\bar{T}_1}}{\partial{\bar{r}}}|_1] ds
 =\frac{\pi}{\epsilon^2} (1+ \frac{k_2}{\kappa_2})
 = \pi (1+ \frac{k_2}{\kappa_2}) V_{\infty}Ma.
\end{equation}
Since both $k_2$ and $\kappa_2$ are positive, we have
\begin{equation}
k_2 \oint \frac{\partial{\bar{T}_2}}{\partial{\bar{r}}}|_1 ds \gg
\oint \frac{\partial{\bar{T}_1}}{\partial{\bar{r}}}|_1 ds.
\end{equation}
From Eqs. (28), we obtain the equivalent integral thermal flux at
the boundary
\begin{equation}
\oint \frac{\partial{\bar{T}_1}}{\partial{\bar{r}}}|_1 ds = k_2
\oint \frac{\partial{\bar{T}_2}}{\partial{\bar{r}}}|_1 ds.
\end{equation}
So, if the thermal flux boundary condition in Eqs.(28) is
satisfied, Eq. (47) should be reduced to Eq. (48), which seems
impossible. It is termed as a nonconservative integral thermal
flux across the surface for the steady thermocapillary droplet
migration at large Ma numbers. This implies the overall
steady-state energy unbalance of two phases in the flow domain in
the co-moving frame of reference.

Eq. (47) indicates that at large Ma numbers the integral thermal
flux across the surface within the droplet is larger than the
surface thermal flux with respect to the continuous phase fluid.
However, it should be noted that the droplet migration is, at that
time, still in an unsteady state. In the analytical and numerical
results\cite{26,27}, the steady migration velocity (in direct
proportion to Ma) is large at high Ma numbers. Under the condition
of large migration velocity, it is unlikely that the order of
 relevant time scale for the transport process to
 generate steady velocity and temperature fields is
 smaller than that for the droplet to move an appreciable distance.
So, due to the variation of physical parameters with the ambient
temperature, a steady migration process may not be reached. Both
experimental results in Fig. 4(a)(b) of \cite{11} and Fig. 4(b)(d)
of \cite{28} clearly display that the thermocapillary droplet
migration at large Ma numbers is in an accelerating state and does
not reach any steady one. Moreover, numerical simulations of the
thermocapillary motion of deformable and non-deformable
 droplets in \cite{18} and the above section 3 indicate that the assumption
 of quasi-steady state is not valid for large Ma numbers. Thus, it is
clear that the invalid assumption of quasi-steady state for the
thermocapillary droplet migration process is a reasonable
explanation for the nonconservative integral thermal flux across
the droplet surface.

\section{Conclusion and discussions}
\label{sec:sum} In this paper, numerical studies are carried out
for thermocapillary migration of a planar non-deformable droplet
in two uniform temperature gradients at moderate and large Ma
numbers by using the front-tracking method. Some calculations at
moderate and large Ma numbers are performed to analyze the
thermocapillary migration for droplets with different sizes. In
the range of droplet radius under study, Re takes moderate values,
and thus the computed flow fields are similar. There appear
different types of migration processes in the time frame under
investigation depending on the values of Ma, which varies within a
large range.
 At moderate Ma numbers, after an increase-decrease process
 in the time evolution of droplet velocity,
the droplet migration reaches a steady state. In the range of
large Ma numbers, the oscillation process in the time evolution of
droplet velocity is transformed into a monotonous accelerating
process as Ma increases. The terminal droplet migration is in an
acceleration process and doesn't reach any steady state. The
instantaneous migration velocity at a fixed migration distance
decreases with increasing Ma number. The numerical simulation
results are in qualitative agreement with experimental ones.

Moreover, in comparing the variations of temperature fields inside
and outside the droplet, it is evident that at large Ma numbers
the weak transport of thermal energy from outside of the droplet
into inside cannot meet the requirement put forward by the steady
migration process, which implies that the advection around the
droplet is a more significant mechanism for heat transfer
across/around the droplet at large Ma numbers.

Furthermore, from the condition of overall steady-state energy
balance in the flow domain, we have identified  a nonconservative
integral thermal flux across the surface for a steady
thermocapillary drop migration in a uniform temperature gradient
at large Ma (Re) numbers. It may well result from the invalid
assumption of quasi-steady state, and this conclusion implies that
the thermocapillary drop migration at large Ma (Re) cannot reach
any steady state and is thus a unsteady process.

We emphasize that all of the above numerical and theoretical
results about the thermocapillary migration system of droplets
involves assumptions of planar non-deformable interfaces and is
subject to constant physical parameters. As mentioned in the
section 1, the simple modelling is easily applied to explore the
dynamical mechanism of droplets, but has potential drawbacks to
reproduce the experimental results\cite{28}. Extension to the more
realistic case of three-dimensional deformable droplets migrating
in a flow field with temperature-dependance physical parameters
remains to be implemented.

\textbf{Acknowledgments} We thank Drs. Z. H. Yin, P. Gao and L.
Chang for discussions and the IMECH/SCCAS SHENTENG 1800/7000
research computing facilities for assisting in the computation.
This work was partially supported by the National Science
Foundation through the Grant No. 11172310.

\newpage
Table I. Physical parameters of the continuous fluid (5cst
Silicone oil) and the droplet (FC-75) at temperature $25^o$C,
which are the working media in the space experiment\cite{28}.

\begin{tabular}{l|llll}
 \hline
        &    $\rho$($g/cm^3$) & $\mu$($10^{-2}dyn s/cm^2$) & $k$($W/mK$) & $\kappa$($10^{-4}cm^2/s$)
\\ \hline
Silicone oil &  0.91      & 4.268 &0.111 & 6.915 \\
FC-75        &  1.77      & 1.416 &0.063 & 2.018 \\
\hline
\end{tabular}

Table II. Correspondence of non-dimensional parameters Re, Ma and
Ca to the droplet radius $R_0$ for the droplet migration in a flow
field with the temperature gradient $G$=12K/cm.

\begin{tabular}{l|lrl}
 \hline
     $R_0(cm)$ &      Re    &   Ma   &  Ca
\\ \hline
0.05         &  0.66        & 44.7    & 0.0044\\
0.15        &   5.93        &  402.5 & 0.013\\
0.25        &   16.5        & 1118.1 & 0.022\\
0.35        &   32.3        & 2191.6 & 0.031\\
0.45        &   53.4        & 3622.8 & 0.040\\
\hline
\end{tabular}

Table III. Correspondence of non-dimensional parameters Re, Ma and
Ca to the droplet radius $R_0$ for the droplet migration in a flow
field with the temperature gradient $G$=9K/cm.

\begin{tabular}{l|lrl}
 \hline
     $R_0(cm)$ &      Re    &   Ma   &  Ca
\\ \hline
0.05        &   0.49        & 33.5    & 0.0033\\
0.15        &   4.45        & 301.9   & 0.010\\
0.25        &   12.4        & 838.6  & 0.017\\
0.35        &   24.2        & 1643.6 & 0.023\\
0.45        &   40.1        & 2717.1 & 0.030\\
\hline
\end{tabular}

\newpage
\textbf{Figure caption}

Fig.~1. Schematic of the computation domain for a planar droplet
migration. The top and bottom walls are non-slip boundaries and
the left and right boundaries are periodic ones.

Fig.~2. Droplet migration velocity versus non-dimensional time for
three grid resolutions $64 \times 128$, $96 \times 192$ and $128
\times 256$ at a fixed domain $4 \times 8$ under Re=5, Ma=20,
Ca=0.01666 and $\rho_2/\rho_1$= $\mu_2/\mu_1$= $k_2/k_1$=
$\kappa_2/\kappa_1$=0.5.

Fig.~3. Time evolution of the droplet migration velocity for grid
resolution $96 \times 192$ and its comparison with Nas \&
Tryggvason's result[15] for the same parameters as given in
Fig.~2.

Fig.~4. Droplet migration velocity in a flow field with the
temperature gradient $G=12$K/cm versus non-dimensional time at
Ma=44.7, 402.5, 1118.1, 2191.6 and 3622.8.

Fig.~5. Instantaneous thermocapillary migration velocity of the
droplet in a flow field with the temperature gradient $G=12$K/cm
at a fixed migration distance $l_z=1/1.5/2cm$ denoted by
diamonds/deltas/squares versus large Ma numbers. The experimental
results\cite{28} rescaled by
$V_{YGB}/v_0=2/[(2+3\mu_2/\mu_1)(2+k_2/k_1)]$ are plotted and
denoted by circles.

Fig.~6. Computed velocity fields at t=20 under $R_0$=0.25cm,
Re=16.5, Ma=1118.1 in (a) the laboratory coordinate frame and (b)
the reference frame moving with the droplet.

Fig.~7. Streamlines in a reference frame moving with the droplet
under $R_0$=0.25cm, Re=16.5, Ma=1118.1. Their time evolution is
displayed in 5 small figures from left to right. The
non-dimensional time is chosen as 3, 10, 20, 40 and 60,
respectively.

Fig.~8. Isotherms in a laboratory coordinate frame are selected
from the computation of the droplet migration under $R_0$=0.05cm,
Re=0.66, Ma=44.7. Notation is the same as in Fig.~7.

Fig.~9. Same as Fig.~8, except $R_0$=0.15cm, Re=5.93, Ma=402.5.

Fig.~10. Same as Fig.~8, except $R_0$=0.35cm, Re=32.3, Ma=2191.6.

Fig.~11. Time evolution of temperature at point $(x_c,z_c)$ inside
the droplet and point $(x_c,z_c+2)$ outside the droplet, where
$(x_c,z_c)$ is the center of the droplet in the laboratory
coordinate system.

Fig.~12. Droplet migration velocity in a flow field with the
temperature gradient $G=9$K/cm versus non-dimensional time at
Ma=44.7, 301.9, 838.6, 1643.7 and 2717.1.

Fig.~13. Instantaneous thermocapillary migration velocity of the
droplet in a flow field with the temperature gradient $G=9$K/cm at
a fixed migration distance $l_z=1/1.5/2cm$ denoted by
diamonds/deltas/squares versus large Ma numbers. The experimental
results\cite{28} rescaled by
$V_{YGB}/v_0=2/[(2+3\mu_2/\mu_1)(2+k_2/k_1)]$ are plotted and
denoted by circles.

\end{document}